\documentclass[preprint,12pt]{elsarticle}

\usepackage{amssymb}

\def\ga{\mathrel{\raise.3ex\hbox{$>$\kern-.75em\lower1ex\hbox{$\sim$}}}}
\def\la{\mathrel{\raise.3ex\hbox{$<$\kern-.75em\lower1ex\hbox{$\sim$}}}}

\journal{Astroparticle Phys., Topical Issue: Cherenkov Telescope Array~~~}

\begin{document}

\begin{frontmatter}




\title{Probes of Lorentz Violation}


\author{John Ellis and Nick E. Mavromatos}

\address{Theoretical Physics and Cosmology Group, Department of Physics, \\
King's College London, London~WC2R 2LS, UK;\\
TH Division, Physics Department, CERN, CH-1211 Geneva 23, Switzerland\\
~~\\
{\rm KCL-PH-TH/2011-33, LCTS/2011-18, CERN-PH-TH/2011-270}}

\begin{abstract}
Lorentz invariance is such an important principle of fundamental physics that it
should constantly be subjected to experimental scrutiny as well as theoretical
questioning. Distant astrophysical sources of energetic photons with rapid time
variations, such as active galactic nuclei (AGNs) and gamma-ray bursters (GRBs),
provide ideal experimental opportunities for testing Lorentz invariance. The
{\v C}erenkov Telescope Array (CTA)
is an excellent experimental tool for making such tests with sensitivities exceeding
those possible using other detectors.

\end{abstract}


\end{frontmatter}


\section{Introduction}
\label{sec:intro}

High-energy astrophysics and related aspects of cosmology are
the bread-and-butter science issues for the {\v C}erenkov Telescope 
Array (CTA)~\cite{cta}, but some jam may be provided by measurements
related to fundamental physics. One such possibility is to use
transient high-energy emissions from distant astrophysical
objects observed by CTA to probe the validity of Lorentz invariance.

From the quantum-mechanical point of view, the vacuum is the
lowest-energy state of a physical system. It should be regarded as
a medium that may have virtual structure, even if it is
devoid of physical particles. As such, it may have non-trivial effects
on particle propagation, even if Lorentz invariance is an underlying
principle. This effect is, of course, familiar in the cases of photons
propagating through plasmas at high temperatures or superconductors
at low temperatures. Might high-energy photons of astrophysical
origin exhibit analogous effects?

This possibility was raised in~\cite{AEMNS}, where it was pointed
out that distant, rapidly-varying astrophysical sources of high-energy
$\gamma$ rays could provide some of the most sensitive probes of some
models of Lorentz invariance. This possibility was raised specifically
in the context of heuristic models of `space-time foam', as
inspired from string theory~\cite{dfoam,mitsou} in particular,
according to
which quantum-gravitational fluctuations in the vacuum could modify
the propagation velocities of photons by amounts that increase with
energy. 

However, similar effects might arise in other theoretical
frameworks~\cite{mestres,loop,dsr,sme,pospelov,lifsh}, and the search for Lorentz violation may be pursued
from a purely phenomenological point of view~\cite{pheno}, prompted by, but not
limited to, specific heuristic models. Lorentz invariance has been one
of the foundations of modern physics for over a century. In the
scientific spirit, it should not be regarded as a sacred principle that
cannot be questioned, but rather as a theoretical dogma that should constantly
be challenged by more sensitive experimental tests. As discussed in this
article, CTA~\cite{cta}  will be uniquely well placed to carry these tests to the next
level of sensitivity, potentially challenging some models of space-time foam.

The structure of this article is as follows. 
Section~\ref{sec:motiv} surveys some physical motivations for the
possibility that energetic photons might travel with speeds smaller
than the classical speed of light, $c$. Section~\ref{sec:current}
reviews the current status of experimental probes of the propagation
speeds of energetic particles, with particular attention to photons but also
including other particles such as electrons and neutrinos.
Section~\ref{sec:prospects} then analyzes the opportunities available
to CTA for extending these probes of Lorentz invariance, notably using
rapidly-varying high-energy $\gamma$ emissions from active galactic
nuclei (AGNs) and gamma-ray bursters (GRBs). For comparison,
Section~\ref{sec:complement} reviews the possible sensitivities of
other future probes of Lorentz invariance, notably those using
astrophysical and terrestrial neutrinos. Finally, Section~\ref{sec:summary}
summarizes the prospects for probing Lorentz violation with CTA.

\section{Motivations}
\label{sec:motiv}

As already mentioned, the idea that the space-time vacuum should
be regarded as a non-trivial medium that may have observable
effects on particles propagating through it~\cite{AEMNS} is a very general one,
much more general than the heuristic models of space-time foam~\cite{dfoam,mitsou}
that spawned the suggestion. Nevertheless, we focus here on the
motivations provided by such models, while mentioning some other
suggestions.

It is a familiar aspect of quantum mechanics that in any physical system
virtual fluctuations with excitation energies $\Delta E$ should arise on
time scales $\Delta t \sim \hslash/\Delta E$. Wheeler extended this
principle to gravity, arguing that quantum-gravitational fluctuations in
the space-time continuum with $\Delta E \sim M_P$ (where $M_P \sim
10^{19}$~GeV is the Planck mass, defined by $1/\sqrt{G_N}$~\cite{Wheeler}, where
$G_N$ is the Newton constant of classical gravity) would endow it with a `foamy'
structure on time scales $\Delta t \sim \hslash/M_P$ (henceforth we use `natural' 
units in which $\hslash, c \equiv 1$). As a result of these quantum-gravitational
fluctuations, Wheeler argued that space-time would no longer appear flat at
distance scales $\Delta x \sim 1/M_P$, possibly with the appearance of
topological fluctuations as well as non-topological irregularities. A lattice
is one example of an inhomogeneous space-time structure, but space-time
foam would presumably have a more irregular, stochastic and aperiodic
structure.

Heuristic models of space-time foam have been proposed, based on
features present in string theory~\cite{dfoam}. This exhibits a plethora of non-perturbative
structures existing in various dimensions, notably D-branes and D-particles~\cite{polchinski}.
A common suggestion in this context is that our Universe is a three-dimension- al
membrane in a higher-dimensional `bulk' space. If D-particles cross `our'
D-brane in this higher-dimensional space, they are perceived in our Universe
as space-time events localized at specific locations $x$ and specific times $t$.

What happens to a photon propagating through our Universe, along `our' D-brane?
In general, if it encounters a D-particle, it will interact with it, much as a photon
propagating through a material medium such as glass may interact with the
electrons that it contains, via absorption and subsequent re-emission. The net 
effect in this case is to slow down the
photon, with the result that light travelling through glass has a positive
refractive index, $\eta$. The value of $\eta$ depends on the colour (i.e., the
wavelength or frequency) of the light, corresponding to the energy of the
associated photon. Analogously, one might expect that light travelling through the
quantum-gravitational vacuum would also acquire an energy-dependent
refractive index, that might be modelled via interactions with D-particles~\cite{emnewuncert}.

This qualitative picture has several important, generic features. The first is
that the refractive index $\eta$ (defined through the photon phase velocity $v_{\rm ph} = p/E = c/\eta$)
should be {\it larger than unity}, corresponding to
{\it subluminal} propagation of energetic photons~\cite{dfoam}. The fact that photons
should not travel faster than $c$ can be argued independently on general grounds: if they
did, they would emit gravitational {\v C}erenkov radiation and lose energy
unacceptably quickly~\cite{MN}. A second generic feature is that the refractive index
should {\it increase} with energy. This is because gravitational interactions
are in general proportional to some power of $G_N$ and increase as some
power of the energy. Simple models~\cite{dfoam,loop,pospelov,dsr} suggest that the
photon group velocity might deviate linearly from that of light:
\begin{equation}
v_g \; \equiv \; \frac{\partial E}{\partial p} \; \sim \; 1 - \frac{E_\gamma}{M_1} , 
\end{equation}
where $M_1$ is some large mass scale that might be ${\cal O}(M_P)$. However,
$M_1$ could depend on other parameters of the microscopic theory, such
as the string scale and/or coupling in a D-particle model~\cite{emnewuncert}, 
as well as the local density of D-particles, and other energy
dependences: $\eta \sim (E/M_n)^n$ should also be considered, in particular the case $n = 2$.

What might happen to the propagation speeds of other types of particles?
The interactions of different particle types with the space-time foam would 
not be universal, in general, so they would have different refractive indices.
In the specific case of the photon, there is no conserved quantum number to impose
a selection rule inhibiting its absorption and re-emission by a space-time 
excitation modelled by a D-particle. On the other hand, the interaction of a 
charged particle such as an electron with these excitations might well be
inhibited~\cite{synchr}. For example, D-particles do not carry electric charges, so in that 
model the electron could not be simply absorbed and re-emitted {\it {\` a} la}
photon, and its refractive index would be suppressed, even vanishing.
A similar argument applies to the proton, with the added complication that
it is a composite particle, further complicating the discussion. The case of
the neutrino is different again: it has no electric charge, and lepton number
is not expected to be absolutely conserved at the scale of quantum
gravity. On the other hand, as a fermion the neutrino could not simply be
absorbed by a space-time defect unless they occur in boson-fermion
doublets {\it {\` a} la} supersymmetry. In any case, the above intuition informs
us that energetic neutrinos should also always propagate {\it subluminally}.

There are many other models of Lorentz violation worthy of mention.
For example, purely phenomenological models have been proposed, motivated by
aspects of cosmic-ray physics~\cite{mestres}. The possibility of spontaneous Lorentz 
violation has been proposed~\cite{sme,pospelov}, and it was argued in some models of loop quantum gravity~\cite{loop}
that the vacuum might exhibit non-trivial optical properties.
Lifshitz-type quantum field theories~\cite{lifsh}, in which the space
and time coordinates have different scaling properties, have been revived
in the context of a new approach to quantum gravity, and have subsequently
attracted more general attention. Lifshitz theories offer a framework in which
Lorentz invariance can be violated fundamentally at high energies, but may be
restored in the low-energy limit. We also note the existence of other theories
with fundamental deformations, rather than violations,  of Lorentz invariance, such as doubly
(or deformed) special relativity~\cite{dsr}. Finally, we recall a number of other
phenomenological approaches~\cite{pheno}.

The existence of this complex ecosystem of Lorentz-violating theories adds
motivation to experimental probes of Lorentz violation, specifically by testing
for a possible deviation $\delta v$ from the speed of light in the propagation of energetic
photons that increases as some power of energy: $\delta v/c \sim - (E/M_n)^n$, in
particular $n = 1$ or 2.

\section{Current Status}
\label{sec:current}

It was proposed in~\cite{AEMNS} that one probe Lorentz violation by searching for the
smearing of transient features in high-energy $\gamma$ emissions from
astrophysical sources. To the extent that such sources are not monochromatic,
the observations at Earth of photons in such emissions would be retarded by 
amounts increasing with energy, and (approximately) coincident arrival times of 
photons with different energies can be used to constrain the parameters $M_n$.
It is clear that the figure of merit for such studies is $L \Delta(E^n)/\Delta t$, where
$L$ is the distance of the source and $\Delta t$ is the time-scale of transient
emissions. Candidate astrophysical sources include pulsars, AGNs and GRBs.

Two cautionary comments are in order. One is that the spectra of transient
emissions from many astrophysical sources become harder with time. Thus, a
higher-energy photon may be relatively more likely to have been emitted later,
potentially polluting any search for a Lorentz-violating refractive index. It is therefore
essential to devise tools for discriminating between retardation effects at source and
during propagation. In principle, this could be done by comparing the smearing of
sources at different distances, since any propagation effect would increase with $L$
whereas (in the absence of evolutionary effects)
source effects should be $L$-independent. However, this strategy entails
the observation of emissions from a number of sources at different distances, and
presupposes the existence of a class of `standard lighthouses' sharing common time
structures in their high-energy emissions. Observations of GRBs are more copious 
than observations of time structures in emissions from AGNs, but neither are very 
uniform `standard lighthouses'. 

The second comment is that an `extraordinary claim requires extraordinary proof'. This 
can be interpreted as implying that any statistical separation between source and 
propagation effects would need to be extraordinarily convincing, but also suggests that 
corroborating evidence from at least two classes of source should be required, e.g., 
both GRBs and AGNs.

The above comments would certainly apply to any claim of an effect. So far, none has
been claimed, and only lower limits on the parameters $M_n$ have been quoted [at least by
experiments using photons ;)]. However, the above comments also apply to claims of
limits. At a minimum, they should be derived from a statistical study of sufficiently many
sources for propagation and source effects to be separated. In the absence of such a
separation, it would be more conservative to regard them as sensitivities to Lorentz
violation. With this caveat in mind, we now review the experimental probes to date,
focusing our attention initially on photons but with some comments later on other
particles such as electrons and neutrinos.

After the initial estimates of sensitivities to Lorentz violation that could be furnished by 
different types of astrophysical source~\cite{AEMNS}, one of the stronger early probes, to
$M_1 \sim 1.8 \times 10^{15}$~GeV, was based on EGRET
observations of the Crab pulsar in the energy range from 70~MeV to 2~GeV~\cite{Kaaret}. By
comparison with AGNs and GRBs, studies of pulsars benefit from the very precise
timing, in the millisecond range, and the possibility of measuring many different pulses.
On the other hand, the prospective sensitivity to Lorentz violation is limited by the
small distances to pulsars, typically $\sim 10^4$~parsec~\cite{pheno}. 
Nevertheless, observations of the Crab pulsar provide
strong limits on Lorentz violation for electrons, as we mention below.
Another early study was based on a TeV $\gamma$-ray flare from the AGN Markarian 421
seen by the Whipple Observatory,
which provided a sensitivity to $M_1 \sim 4 \times 10^{16}$~GeV~\cite{Biller}. However, this first AGN
study had relatively low statistics, and neither of these pioneering studies was able to
disentangle source and propagation effects.

The first study to attempt this by combining data from several different sources at different 
distances was made using a sample of GRB emissions observed by BATSE and OSSE~\cite{mitsou}, 
which gave a sensitivity to $M_1 \sim 10^{15}$~GeV. The use of nine GRBs
with measured redshifts between $z = 0.0085$ and 3.9 laid the basis for a more robust analysis
that was sensitive to $M_1 \sim 6.9 \times 10^{15}$~GeV~\cite{robust},
but the fact that GRB emissions are very irregular, with no obvious `standard lighthouse'
characteristics, complicated the disentanglement of source effects. This analysis was
subsequently upgraded to a fit analyzing emissions from 35 GRBs with measured
redshifts $z \le 6.29$ observed by the BATSE, HETE and SWIFT instruments, which yielded the robust lower
limit $M_1 \ga 1.4 \times 10^{16}$~GeV at the 95\% CL~\cite{morerobust}, despite a strong correlation between the
source and propagation parameters in the fit.

\begin{figure}[ht]
\centering
\includegraphics[width=5cm,angle=-90]{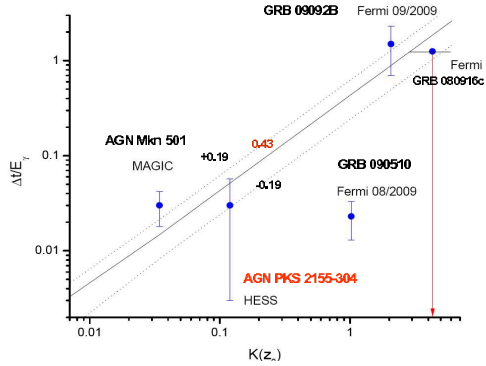}
\caption{\it Comparison of data on delays
$\Delta t$ in the the arrival times of energetic gamma rays from various astrophysical sources with
models in which the velocity of light is reduced by an amount linear in the photon
energy~\protect\cite{Dvoid}. The graph plots on a logarithmic scale the quantity $\Delta t/E$ and a function of the
red-shift, $K(z)$, which is essentially the distance of the source from the observation point.
The data include two AGNs, Mkn 501~\protect\cite{MAGIC} and PKS
2155-304~\protect\cite{HESS}, and three GRBs observed by the Fermi satellite,
090510, 09092B and 080916c~\protect\cite{Fermi}.}%
\label{fig:data}%
\end{figure}

Fig.~\ref{fig:data} compares the sensitivities of different experiments to the time-lag/energy ratio
$\Delta t/E$ as functions of the quantity $K(z)$ defined by
\begin{equation}
K (z)  \; \equiv \; \frac{1}{(1 + {\hat z})} \int_0^z \frac{(1 + {\hat z}) {\rm d {\hat z}}}{h({\hat z})} ,
\label{K}
\end{equation}
where the Hubble expansion rate $h({\hat z}) = \sqrt{\Omega_\Lambda + \Omega_M (1 + {\hat z})^3}$,
which measures the distance of the source from the observation point.
As seen in Fig.~\ref{fig:data}, greater sensitivities to $\Delta t/E$, and hence $M_1$,
have been found more recently in in two analyses of AGN flares.
The first was a flare of $\gamma$-rays with energies $\la 10$~TeV from Markarian 501 
at $z = 0.034$ observed by the MAGIC telescope, which had sensitivity to
$M_1 \sim 2 \times 10^{17}$~GeV~\cite{MAGIC}. The second was a flare of $\gamma$-rays with energies
$\la 1$~TeV from PKS 2155-304 at redshift $z = 0.116$ observed by the H.E.S.S. telescope,
which yielded sensitivity to $M_1 \sim 2 \times 10^{18}$~GeV~\cite{HESS}. Several factors contributed
to the greater sensitivity of this observation: the higher redshift, the higher range of $\gamma$
energies and the larger statistics. All of these are features that CTA may hope to improve
further, principally by virtue of its much greater collection power.

The principal competition for these AGN observations is currently provided by Fermi GBM/LAT
measurements of energetic $\gamma$-rays from GRBs. As also seen in Fig.~\ref{fig:data}, the most sensitive of these to date
has been provided by observations of GRB 090510 at $z = 0.90$, in particular the observation of a
single photon with $E_\gamma \sim 31$~GeV. The sensitivity to $M_1$ inferred from this single
photon depends whether one assumes (most conservatively) that it was not emitted before the
start of any detectable emission below 1~MeV, $M_1 \sim 1.4 \times 10^{19}$~GeV, or that
it was not emitted before the start of emission above 1~GeV, $M_1 \sim 1.2 \times 10^{20}$~GeV, 
or (most aggressively) that it was associated with the nearest 1~MeV emission spike, in which case
the sensitivity is to $M_1 \sim 1.2 \times 10^{21}$~GeV~\cite{Fermi}. This remarkable sensitivity is traceable
in part to the cosmological redshift and the fine time structure, which are typical of GRB emissions,
as well as to the observation of the highest-energy $\gamma$-ray seen from a GRB so far.

The previous paragraphs focused on the possibility of a photon speed decreasing linearly with energy,
where the finer time structure of GRB emissions gives them an advantage over AGN emissions.
However, the boot is on the other foot when one considers a quadratic dependence: $\delta v \sim - (E/M_2)^2$.
In this case, the higher energies seen in AGN emissions confer on them a significant advantage. For
example, the MAGIC data on Markarian 501 yield a sensitivity to $M_2 \sim 2.6 \times 10^{10}$~GeV~\cite{MAGIC}
and those on PKS 2155-304 are sensitive to $M_2 \sim 6.4 \times 10^{10}$~GeV~\cite{HESS},
with which the observations of GRB 090510~\cite{Fermi} are not competitive.

As mentioned in Section~\ref{sec:motiv}, Lorentz-violating effects on the propagation of
different particle species are not expected to be universal, and sensitivities to (constraints on)
Lorentz violation for electrons, protons and neutrinos are of independent interest in their own
rights. We note, in particular, that broadband observations of electromagnetic radiation from
the Crab Nebula restrict very severely any possible Lorentz violation for the electron, so that
the corresponding $M_1^e \gg M_P$~\cite{pheno}. Important constraints on Lorentz violation for other particles are
provided by observations of ultra-high-energy cosmic rays (UHECRs), via constraints on both
conventional processes such as $p + \gamma \to p + \pi^0, n + \pi^+$ and unconventional
processes such as $p \to p + \gamma, p + \pi^0$ that are forbidden by Lorentz invariance~\cite{UHECRs}~\footnote{There
are also very stringent constraints~\cite{pheno} on models that predict birefringence~\cite{pospelov}, 
unlike the stringy models of \cite{dfoam}.}.

There has been much interest recently in the possibility of probing Lorentz violation in the
propagation of neutrinos. The rough coincidence of the neutrino pulse from the supernova SN1987a
with the onset of optical emissions imposed $|\delta v| \la 10^{-9}$ at neutrino energies 
$\sim 10$~MeV~\cite{SNfixed}, much stronger than limits at GeV energies established by MINOS: 
$\delta v = (5.1 \pm 2.9) \times 10^{-5}$~\cite{MINOS} [superluminal by $1.8 \sigma$], and
earlier accelerator neutrino experiments. 
The fact that the observed SN1987a neutrino pulse was not broadened relative to the
predictions of a core-collapse supernova model was used in~\cite{Harries} to establish upper 
limits on the possible energy dependence of $\delta v$: $M_1^\nu > 2.5 \times 10^{10}$~GeV
and $M_2^\nu > 4.1 \times 10^4$~GeV for superluminal propagation, and similar limits for
the subluminal case~\cite{Harries}. The same paper
suggested how the OPERA experiment~\cite{OPERAExp} could be used to improve on the previous accelerator
neutrino constraint by exploiting the time structure of the CERN CNGS neutrino beam~\cite{Harries}.

The OPERA experiment has recently reported evidence for {\it superluminal} neutrino
propagation: $\delta v \sim + 2.5 \times 10^{-5}$ for neutrinos with $\langle E_\nu \rangle \sim 28$~GeV~\cite{OPERA}. 
If the amount of Lorentz violation were linear (quadratic) in the neutrino energy, this would correspond to
$M_1^\nu \sim 1.1 \times 10^6$~GeV ($M_2^\nu \sim 5.6 \times 10^3$~GeV), in conflict with the SN1987a constraint~\cite{AEM,caccia}. 
Such a large value of $\delta v$ is also 
constrained by limits on both electromagnetic~\cite{CG} and gravitational {\v C}erenkov radiation~\cite{AEM,MN}. 

At the time of writing, the OPERA experiment is making a new measurement using a modified CNGS beam,
and it is uncertain whether the OPERA evidence will survive further experimental scrutiny. 
We emphasize that the present and prospective constraints on Lorentz violation using photons are
logically independent of those using neutrinos, as well as being much more stringent. Nevertheless,
the OPERA report has reminded the physics community of the importance of increasing the
sensitivity of experimental probes of Lorentz violation.

\section{Prospects for CTA}
\label{sec:prospects}

The large collection area and acceptance of CTA will provide it with unique prospects for gathering data on
astrophysical sources of energetic $\gamma$-rays, and hence new opportunities for probing Lorentz
violation. 

On the one hand, CTA will be able to make much more detailed studies of known sources such 
as the AGNs Markarian 421, Markarian 501 and PKS 2155-304, which should provide higher-energy
$\gamma$-rays as well as greater statistics, and might reveal smaller structures
in their $\gamma$ emissions. Clearly observations of higher-energy $\gamma$-rays from
astrophysical objects at similar redshifts improve the sensitivity to Lorentz violation. In addition,
greater statistics in the observations of structures in the
emissions from specific objects will improve the understanding of their source effects. Moreover,
CTA should be able to observe structures with smaller amplitudes than those accessible to previous
$\gamma$-ray telescopes. Such small-amplitude structures might be associated with emissions
from smaller regions of the AGNs, that therefore could exhibit shorter time-scales and provide improved
sensitivity to Lorentz violation.

On the other hand, CTA will also be able to extend $\gamma$-ray observations to a larger family of
less-luminous AGNs. These will include objects with similar intrinsic luminosities to known AGNs, 
but with larger redshifts, improving the
sensitivity to Lorentz violation and providing a longer lever arm for separating source and
propagation effects, as well as making
possible more systematic studies of the source effects themselves. CTA may also detect a population
of nearby objects with less intrinsic luminosity. {\it A priori}, these objects might have smaller emission regions
and hence exhibit more rapid flux variations, yielding enhanced sensitivity to Lorentz violation.

The improvements in sensitivity to Lorentz violation that will be provided by CTA are
somewhat uncertain, being hostage to the stochastic properties of $\gamma$-ray emissions.
However, based on the above discussion one can consider three generic possibilities: higher energies, larger redshifts and
finer time structures.

A typical spectral index $\Gamma: dN/dE_\gamma \sim (E_\gamma/E_0)^{- \Gamma}$ 
for AGN $\gamma$-ray emissions is $\Gamma \sim 2$~\cite{HESS},
implying, e.g., that a collecting power $\sim 100$ greater than that of H.E.S.S. would be required to
extend its observations of PKS~2155-304 to $\gamma$-rays with $\langle E_\gamma \rangle \sim 10$~TeV.
Nevertheless, if this could be achieved, and assuming the observation of a transient structure in emissions
with a similar time-scale, it would extend the sensitivity reported in~\cite{HESS} to
$M_1 \sim 10^{19}$~GeV, tantalizingly close to the Planck scale.

\begin{figure}
\centering
\includegraphics[width=6cm]{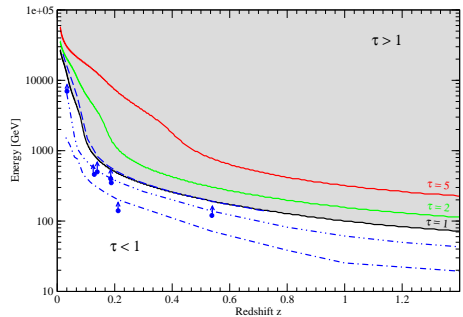}
\caption{\it The horizon for high-energy photons interacting with 
the cosmic infrared and microwave backgrounds, shown as solid lines corresponding to various
optical depths $\tau$~\protect\cite{KD}. The dashed and dot-dashed curves
correspond to alternative models of the extragalactic background light.}
\label{fig:PM}
\end{figure}

However, there is a potential limitation to what can be achieved with higher-energy $\gamma$-rays, since
their mean free path for interaction with (and hence energy loss to) microwave and infrared
background photons imposes an effective limit on the energies of photons emitted at large redshifts that 
can in fact be observed. For example, it is calculated that interactions with the cosmic microwave background would prevent
the observations of photons with energies $E_\gamma > 100$~TeV emitted from more than a few Mpc away.
The cosmic infrared background is less well determined but, as seen in Fig.~\ref{fig:PM},
it is likely to absorb photons with $E_\gamma > 1$~TeV
emitted by AGNs at distances larger than those of Markarian 421, Markarian 501 and PKS 2155-304~\cite{KD}. This was
highlighted in~\cite{PM} as an issue for the observation of $\sim 20$-TeV $\gamma$-rays from Markarian 501 
reported by HEGRA~\cite{HEGRA}. It was also noted in~\cite{PM} that Lorentz violation 
might actually resolve this problem, though more mundane explanations are also possible~\cite{Dwek}.

The same issue arises in considering the potential for observing emissions from larger redshifts. For
example, it is seen in~\cite{KD}, assuming conventional Lorentz-invariant kinematics, that only
photons with $E_\gamma \la 300$~GeV are likely to be observable in emissions from $z \sim 1$.
Nevertheless, if one assumes that a structure similar to that observed by H.E.S.S. in emissions from
PKS 2155-304 were to to be observed in TeV-scale emissions from an AGN with $z \sim 1$, which might
be possible with the collecting power of CTA, this would also give sensitivity to $M_1 \sim 10^{19}$~GeV.

Alternatively, one may consider the possibility of observing transient emissions with shorter time-scales
than those observed from AGNs so far. These time-scales are presumably related to the sizes of
accretion regions in the AGNs, which are, in turn, loosely related to the overall sizes of the cores of the
AGN. Shorter time-scales might arise from AGNs with smaller cores, which are likely to be less
luminous. Or they might arise from `hotspots' of accretion corresponding to small portions of the overall
emission region. In either case, the magnitudes of any small-time-scale fluctuations are likely also to be
of smaller amplitude than those observed from PKS 2155-304 observed by H.E.S.S., for example. 
However, if the larger statistics obtainable with CTA were to reveal structures with time-scales an order
of magnitude shorter than that observed from PKS 2155-304, but with similar energies and from similar 
redshifts, sensitivity to $M_1 \sim 10^{19}$~GeV might again be attained.

We therefore conclude that there are various ways in which CTA could offer sensitivity to $M_1 \sim M_P$,
the characteristic mass-scale of quantum gravity.

\section{Complementarity with other experiments}
\label{sec:complement}

The principal competition for CTA in probing Lorentz violation with photons will be
provided by experiments studying emissions from GRBs. As we have seen, Fermi GBM/LAT
observations of GRB 090510 at $z = 0.90$ already provide sensitivity to $M_1 \sim M_P$,
and potentially greater sensitivity if one makes more aggressive analysis assumptions~\cite{Fermi}.
However, so far this sensitivity is provided by a single photon that is considerably more
energetic than others emitted by GRBs. On the other hand, GRB 090510 did not have a
particularly large redshift, the largest known having been GRB 090423 with a redshift $z = 8.2$.
However, since a refractive index $\eta \sim E_\gamma/M_1$ yields a time delay proportional
to $K(z)$ (\ref{K}), which is significantly nonlinear at large redshifts,
there is no great advantage in sources with $z > 1$. Rather, consolidation of the sensitivity attained
with GRB 090510 will require significantly more statistics of comparably energetic photons,
which are currently in short supply with the Fermi satellite. Moreover, establishment of a solid
limit on $M_1$ (either lower or upper) would require concordant measurements with different classes of
sources, with their different source effects, e.g., both GRBs and AGNs.

It should also be emphasized that CTA will come into its own when the sensitivity to a possible
quadratic dependence of the speed on energy is considered, in view of its ability to measure
photons of much higher energies than those typical of GRB emissions.

Another point worth noting is that, in at least some models of space-time, the refractive index
may vary with redshift. This is because, e.g., in the D-particle model discussed earlier the
density of D-particles traversing the three-dimensional brane constituting our Universe may
well depend on the cosmological epoch, i.e., $z$~\cite{Dvoid}. In order to avoid this ambiguity, one should
compare with measurements of AGNs those of emissions from GRBs with similar redshifts
$z = {\cal O}(0.1)$. The sensitivity to Lorentz violation provided by GRBs in this $z$ range is
currently significantly less than that attained with AGN measurements.

As we have already discussed, there is no direct comparison between photon probes of
Lorentz violation and probes using other particles. Specifically, we would not expect similar
effects for charged particles such as electrons and protons, though perhaps for neutrins. 
As we have also seen above, probes of Lorentz violation using neutrinos
currently have many orders of magnitude less sensitivity than those using photons. However,
there are significant possibilities for improvement. For example, it has recently been found in
two-dimensional simulations of core-collapse supernovae that their neutrino emissions may
exhibit time structures on the millisecond scale~\cite{2D}. If observed, these structures would immediately
improve the sensitivity to $M_1^\nu$ by two orders of magnitude~\cite{Janka}, though still much
less than for photons.

There are also prospects for increasing the sensitivity to $M_1^\nu$ of long-baseline
accelerator neutrino experiments. As discussed in~\cite{Harries}, it should be possible
to improve the CNGS timing measurement to the level of 1~ns, offering an improvement
in sensitivity by an order of magnitude. It is in principle possible to improve the time
resolution of particle detectors to the level of ${\cal O}(1)$~ps, though how to exploit this in
a long-baseline experiment is an open question. In any case, for the foreseeable future
neutrino experiments will have less sensitivity than photon experiments such as CTA, and
are in any case measuring an independent quantity.

\section{Summary}
\label{sec:summary}

As we have reviewed in this article, there is considerable interest in probing the possible
violation of Lorentz invariance: precisely because it is so fundamental to modern physics,
it should not be treated as sacrosanct, but rather as a principle to be questioned assiduously.
As we have discussed, distant astrophysical objects such as GRBs and AGNs offer the most
sensitive opportunities to probe Lorentz violation using energetic photons. Probes using
accelerator and astrophysical neutrinos are much less sensitive, and the very strong limits 
on Lorentz violation obtained using electrons do not have immediate implications for photons.

As we have also discussed, CTA will have a unique scientific opportunity to probe Lorentz
violation using astrophysical photons, in particular multi-TeV $\gamma$ rays from AGNs. The
sensitivity that can be attained by CTA probes is subject to the unknown vagaries of energetic
astrophysical sources, and is restricted by the horizon for absorption of energetic
$\gamma$ rays by extragalactic background light. However, these restrictions can be evaded
by observing emissions with short transient time-scales, so the sky is not quite the limit for CTA
to probe Lorentz violation.

\section*{Acknowledgements}
This work was supported in part by the London
Centre for Terauniverse Studies (LCTS), using funding from the European
Research Council 
via the Advanced Investigator Grant 267352. 






\begin{thebibliography}{00}

\bibitem{cta}
  G.~Hermann [CTA Collaboration],
  Nucl.\ Phys.\ Proc.\ Suppl.\  {\bf 212-213 } (2011)  170-177 and references therein.
  

\bibitem{AEMNS}
  G.~Amelino-Camelia, J.~R.~Ellis, N.~E.~Mavromatos, D.~V.~Nanopoulos, S.~Sarkar,
  Nature {\bf 393 } (1998)  763-765
  [astro-ph/9712103].
  
  
\bibitem{dfoam} G.~Amelino-Camelia, J.~R.~Ellis, N.~E.~Mavromatos, D.~V.~Nanopoulos,
  Int.\ J.\ Mod.\ Phys.\  {\bf A12 } (1997)  607-624
  [hep-th/9605211]; 
J.~R.~Ellis, N.~E.~Mavromatos, D.~V.~Nanopoulos,
  Gen.\ Rel.\ Grav.\  {\bf 32 } (2000)  127-144
  [gr-qc/9904068];
J.~R.~Ellis, N.~E~Mavromatos, M.~Westmuckett,
  Phys.\ Rev.\  {\bf D70 } (2004)  044036
  [gr-qc/0405066]; 
  \emph{ibid.} {\bf D71 } (2005)  106006
  [gr-qc/0501060];
   J.~R.~Ellis, N.~E.~Mavromatos, D.~V.~Nanopoulos, M.~Westmuckett,
  Int.\ J.\ Mod.\ Phys.\  {\bf A21 } (2006)  1379-1444
  [gr-qc/0508105].  
  
   \bibitem{mitsou}  J.~R.~Ellis, K.~Farakos, N.~E.~Mavromatos, V.~A.~Mitsou, D.~V.~Nanopoulos,
  Astrophys.\ J.\  {\bf 535 } (2000)  139-151
  [astro-ph/9907340]. 
  
  
      
  \bibitem{mestres} L.~Gonzalez-Mestres,
  arXiv:physics/9712005 and \\
    arXiv:physics/9708028.  
  
  
    
    \bibitem{loop} R.~Gambini, J.~Pullin,
  Phys.\ Rev.\  {\bf D59 } (1999)  124021
  [gr-qc/9809038].

    \bibitem{dsr} G.~Amelino-Camelia,
  Int.\ J.\ Mod.\ Phys.\  {\bf D11 } (2002)  35-60
  [gr-qc/0012051];
  J.~Magueijo, L.~Smolin,
  Phys.\ Rev.\ Lett.\  {\bf 88 } (2002)  190403
  [hep-th/0112090].
  
\bibitem{sme}
  V.~A.~Kostelecky,
  arXiv:hep-ph/9912528.  
  
\bibitem{pospelov} 
  R.~C.~Myers, M.~Pospelov,
  Phys.\ Rev.\ Lett.\  {\bf 90 } (2003)  211601
  [hep-ph/0301124];
  J.~Alfaro, L.~F.~Urrutia,
  Phys.\ Rev.\  {\bf D81 } (2010)  025007
  [arXiv:0912.3053 [hep-ph]].  
  
  \bibitem{lifsh} 
  P.~Horava,
  Phys.\ Rev.\  {\bf D79 } (2009)  084008
  [arXiv:0901.3775 [hep-th]];
  M.~Visser,
  Phys.\ Rev.\  {\bf D80}, 025011 (2009)
  [arXiv:0902.0590 [hep-th]];
  for a comprehensive review in our context, see: 
  J.~Alexandre,
    arXiv:1109.5629 [hep-ph] and references therein.
  
 \bibitem{pheno} 
 S.~R.~Coleman, S.~L.~Glashow,
  Phys.\ Rev.\  {\bf D59 } (1999)  116008
  [hep-ph/9812418];
  T.~Jacobson, S.~Liberati, D.~Mattingly,
  Nature {\bf 424 } (2003)  1019-1021
  [astro-ph/0212190];
  Annals Phys.\  {\bf 321 } (2006)  150-196
  [astro-ph/0505267]; 
  N.~E.~Mavromatos,
  Int.\ J.\ Mod.\ Phys.\  A {\bf 25} (2010) 5409
  [arXiv:1010.5354 [hep-th]];
  S.~Liberati, L.~Maccione,
  J.\ Phys.\ Conf.\ Ser.\  {\bf 314 } (2011)  012007
  [arXiv:1105.6234 [astro-ph.HE]] and references therein.  
  
  
  \bibitem{Wheeler}
 See, for example,  J.~A.~Wheeler and K.~W.~Ford, \emph{Geons, black holes, and quantum foam : a life in physics}
 (Norton, New York, 1998) ISBN 0-393-04642-7.
  
\bibitem{polchinski}
  J.~Polchinski,
  Phys.\ Rev.\ Lett.\  {\bf 75 } (1995)  4724-4727
  [hep-th/9510017] and
  \emph{String theory. Vol. 2: Superstring theory and beyond},
  (Cambridge Univ. Press, Cambridge, UK, 1998).
    
    
\bibitem{emnewuncert}
  J.~R.~Ellis, N.~E.~Mavromatos, D.~V.~Nanopoulos,
  Phys.\ Lett.\  {\bf B665 } (2008)  412-417
  [arXiv:0804.3566 [hep-th]];
  Int.\ J.\ Mod.\ Phys.\  {\bf A26 } (2011)  2243-2262
  [arXiv:0912.3428 [astro-ph.CO]];
  T.~Li, N.~E.~Mavromatos, D.~V.~Nanopoulos, D.~Xie,
  Phys.\ Lett.\  {\bf B679 } (2009)  407-413
  [arXiv:0903.1303 [hep-th]].


\bibitem{MN}
G.~D.~Moore and A.~E.~Nelson,
  JHEP {\bf 0109} (2001) 023
  [arXiv:hep-ph/0106220].
  
\bibitem{synchr}   J.~R.~Ellis, N.~E.~Mavromatos, A.~S.~Sakharov,
  Astropart.\ Phys.\  {\bf 20 } (2004)  669-682;
  [astro-ph/0308403].
  J.~R.~Ellis, N.~E.~Mavromatos, D.~V.~Nanopoulos, A.~S.~Sakharov,
  Int.\ J.\ Mod.\ Phys.\  {\bf A19 } (2004)  4413-4430
  [gr-qc/0312044].
  
  
  
\bibitem{Kaaret}
  P.~Kaaret,
  arXiv:astro-ph/9903464.

    \bibitem{Biller} 
  S.~D.~Biller {\it et al.},
  Phys.\ Rev.\ Lett.\  {\bf 83} (1999) 2108
  [arXiv:gr-qc/9810044].

    
 \bibitem{robust} 
  J.~R.~Ellis, N.~E.~Mavromatos, D.~V.~Nanopoulos, A.~S.~Sakharov,
  Astron.\ Astrophys.\  {\bf 402 } (2003)  409-424;
  [astro-ph/0210124].
  
  \bibitem{morerobust}
  J.~R.~Ellis, N.~E.~Mavromatos, D.~V.~Nanopoulos, A.~S.~Sakharov and E.~K.~G.~Sarkisyan,
  Astropart.\ Phys.\  {\bf 25} (2006) 402
  [Astropart.\ Phys.\  {\bf 29} (2008) 158](E)
  [arXiv:astro-ph/0510172].
  
  
\bibitem{Dvoid} 
  J.~R.~Ellis, N.~E.~Mavromatos, D.~V.~Nanopoulos,  
  Int.\ J.\ Mod.\ Phys.\  {\bf A26 } (2011)  2243-2262
  [arXiv:0912.3428 [astro-ph.CO]].


  
  
\bibitem{MAGIC} 
  J.~Albert {\it et al.} [MAGIC Collaboration], J.~Ellis, N.~E.~Mavromatos, D.~V.~Nanopoulos, A.~S.~Sakharov and E.~K.~G.~Sarkisyan,
  Phys.\ Lett.\  {\bf B668 } (2008)  253-257
  [arXiv:0708.2889 [astro-ph]].

 \bibitem{HESS}
A.~Abramowski {\it et al.}  [HESS Collaboration],
  Astropart.\ Phys.\  {\bf 34} (2011) 738
  [arXiv:1101.3650 [astro-ph.HE]].

\bibitem{Fermi}   M.~Ackermann {\it et al.} [Fermi GBM/LAT Collaboration],
  Nature {\bf 462 } (2009)  331-334
  [arXiv:0908.1832 [astro-ph.HE]];
  see also: 
 F.~de Palma [Fermi-LAT and Fermi GBM Collaborations],
  AIP Conf.\ Proc.\  {\bf 1223 } (2010)  173-182 and references therein.


\bibitem{UHECRs}
  M.~Galaverni, G.~Sigl,
  Phys.\ Rev.\ Lett.\  {\bf 100 } (2008)  021102
  [arXiv:0708.1737 [astro-ph]];
  Phys.\ Rev.\  {\bf D78 } (2008)  063003
  [arXiv:0807.1210 [astro-ph]];
  L.~Maccione, S.~Liberati, G.~Sigl,
  Phys.\ Rev.\ Lett.\  {\bf 105 } (2010)  021101.
  [arXiv:1003.5468 [astro-ph.HE]].




\bibitem{SNfixed}  K.~Hirata {\it et al.} [KAMIOKANDE-II Collaboration],
  Phys.\ Rev.\ Lett.\  {\bf 58 } (1987)  1490-1493;
  R.~M.~Bionta, {\it et al.} [IMB Collaboration],
  Phys.\ Rev.\ Lett.\  {\bf 58 } (1987)  1494;
   E.~N.~Alekseev, L.~N.~Alekseeva, I.~V.~Krivosheina and V.~I.~Volchenko,
  Phys.\ Lett.\  {\bf B205} (1988) 209.
   M.~J.~Longo,
  Phys.\ Rev.\  {\bf D36 } (1987)  3276;
  L.~Stodolsky, Phys.\ Lett.\  {\bf B201} 353 (1988).
  
\bibitem{MINOS}
  P.~Adamson {\it et al.} [MINOS Collaboration],
  Phys.\ Rev.\  {\bf D76 } (2007)  072005
  [arXiv:0706.0437 [hep-ex]];
  see also
  G.~R.~Kalbfleisch, N.~Baggett, E.~C.~Fowler and J.~Alspector,
  Phys.\ Rev.\ Lett.\  {\bf 43} (1979) 1361.
  

\bibitem{Harries}  J.~R.~Ellis, N.~Harries, A.~Meregaglia, A.~Rubbia, A.~Sakharov,
  Phys.\ Rev.\  {\bf D78 } (2008)  033013.
  [arXiv:0805.0253 [hep-ph]].


\bibitem{OPERAExp}  R.~Acquafredda {\it et al.} [OPERA Collaboration],
  JINST {\bf 4 } (2009)  P04018
  
\bibitem{OPERA}
  T.~Adam {\it et al.} [OPERA Collaboration],
    [arXiv:1109.4897 [hep-ex]].

\bibitem{AEM}
  J.~Alexandre, J.~Ellis, N.~E.~Mavromatos,
    [arXiv:1109.6296 [hep-ph]]

\bibitem{caccia}
  G.~Cacciapaglia, A.~Deandrea, L.~Panizzi,
    [arXiv:1109.4980 [hep-ph]]; 
  G.~F.~Giudice, S.~Sibiryakov, A.~Strumia,
    [arXiv:1109.5682 [hep-ph]].    


\bibitem{CG}   A.~G.~Cohen, S.~L.~Glashow,
  [arXiv:1109.6562 [hep-ph]].

\bibitem{KD}
T.~M.~Kneiske and H.~Dole,
  arXiv:1001.2132 [astro-ph.CO];
  see also D.~Mazin {\it et al.}, for the CTA Collaboration, {\it Potential of EBL and cosmology studies with the
  Cherenkov Telescope Array} (2011).
  
\bibitem{PM}
  R.~J.~Protheroe, H.~Meyer,
  Phys.\ Lett.\  {\bf B493 } (2000)  1-6
  [astro-ph/0005349].
  
  
\bibitem{HEGRA}
  F.~Aharonian {\it et al.} [HEGRA Collaboration],
  Astrophys.\ J.\  {\bf 546 } (2001)  898-902
  [astro-ph/0008211].

\bibitem{Dwek}
  E.~Dwek and F.~Krennrich,
  Astrophys.\ J.\  {\bf 618} (2005) 657
  [arXiv:astro-ph/0406565];
  Astrophys.\ J.\  {\bf 733} (2011) 77
  [arXiv:1101.3498 [astro-ph.CO]].
  
  
\bibitem{2D}
  R.~Buras, H.-T.~Janka, M.~Rampp and K.~Kifonidis,
  Astron. \& Astrophys. {\bf 457}, 281 (2006)
  [arXiv:astro-ph/0512189];
  A.~Marek and H.-T. Janka,
  Astrophys.\ J.\ {\bf 694}, 664 (2009)
  [arXiv:0808.4136 [astro-ph]].
   
\bibitem{Janka}
  J.~Ellis, H.~T.~Janka, N.~E.~Mavromatos, A.~S.~Sakharov and E.~K.~G.~Sarkisyan,
  arXiv:1110.4848 [hep-ph].
    
  


\end{thebibliography}



\end{document}